\documentclass[10pt,twocolumn,english,jcp, aip]{revtex4-1}
\usepackage[T1]{fontenc}
\usepackage[latin9]{inputenc}
\usepackage{amsmath}
\usepackage{amssymb}
\usepackage{graphicx}
\usepackage{esint}

\makeatletter
\@ifundefined{date}{}{\date{}}
\usepackage{babel}

\makeatother

\usepackage{babel}
\begin{document}

\title{Note: On the memory kernel and the reduced system propagator }

\author{Lyran Kidon}

\affiliation{\small{Department of Chemistry, University of California, Berkeley, California 94720USA}}

\affiliation{\small{Materials Science Division, Lawrence Berkeley National Laboratory, Berkeley, California 94720, USA}}

\author{Haobin Wang}

\affiliation{\small{Department of Chemistry, University of Colorado Denver, Denver, Colorado 80217-3364, USA}}

\author{Michael Thoss}

\affiliation{\small{Institute of Physics, University of Freiburg, Hermann-Herder-Strasse 3, 79104 Freiburg, Germany}}

\author{Eran Rabani}

\affiliation{\small{Department of Chemistry, University of California, Berkeley, California 94720USA}}

\affiliation{\small{Materials Science Division, Lawrence Berkeley National Laboratory, Berkeley, California 94720, USA}}

\affiliation{\small{The Sackler Center for Computational Molecular and Materials Science, Tel Aviv University, Tel Aviv, Israel 69978}}

\maketitle
The generalized quantum master equation (GQME) formalism has recently
been proven highly successful for the calculation of the reduced dynamics
of complex many-body systems driven away from equilibrium. Two approaches
exist: the Nakajima--Zwanzig--Mori time-convolution (TC) approach,\citep{nakajima_quantum_1958,zwanzig_ensemble_1960,mori_transport_1965}
and the Tokuyama--Mori time-convolutionless (TCL) approach.\citep{Tokuyama1976statistical}
In both approaches, the complexity of solving the many-body quantum
Liouville equation is replaced with the need to evaluate a time-dependent
kernel/super-operator from which the dynamics of the reduced system
can be inferred at all times. The formalism becomes advantageous when
the characteristic decay time of the kernels is short compared to
the approach to equilibrium or steady state,\citep{Geva2003,cohen_memory_2011,Cohen2013kondo,wilner_bistability_2013,wilner2014,Kidon2015TCL}
whereby brute-force numerically-converged solvers~\citep{Makri1995,muhlbacher_real-time_2008,weiss_iterative_2008,Wang2009,gull10_bold_monte_carlo,Segal10,Cohen-Gull-Reichman2015-introducing-inchworm}
provide efficient schemes to obtain the kernels. Alternatively, the
GQME is also an excellent starting point for approximate schemes based
on semiclassical and mixed quantum-classical approaches.\citep{zhang_nonequilibrium_2006,Berkelbach2012,Montoya-Castillo-and-Markland2016when_can_one_win,Montoya-Castillo-and-Reichman2016-nonequilibrium-Nakajima-Zwanzig}

The TC memory kernel and the TCL kernel/generator are computed by
means of auxiliary super-operators. For the TC memory kernel, this
involves the solution of an integro-differential equation and requires
the calculation of a super-operator whose form depends explicitly
on the system-bath coupling,\citep{Geva2003,zhang_nonequilibrium_2006}
which often becomes an intractable task. For the TCL kernel, the calculation
is based on the so-called reduced system propagator\emph{ }super-operator
which requires only generic system observables,\citep{Laird1991,Kidon2015TCL}
but requires an inversion of the reduced propagator which can lead
to numerical instabilities.

Here, we outline a simple framework to obtain the TC memory kernel
from the reduced system propagator alone, circumventing the need to
obtain its inverse (TCL) or calculate higher order system-bath observables
(TC). The form of the reduced system propagator is \emph{universal}
for any system-bath Hamiltonian, allowing for a reduction in the complexity
of obtaining the TC memory kernel. We demonstrate this on the nonequilibrium
generalized Anderson-Holstein impurity model. The framework also provides
direct relations between the TC and TCL kernels in terms of the reduced
system propagator.

Consider an open quantum system coupled to an environment (bath),
described by ${\cal H}={\cal H}_{S}+{\cal H}_{B}+{\cal H}_{SB}$,
where ${\cal H}_{S}$,~${\cal H}_{B}$ are the system and bath Hamiltonians,
and ${\cal H}_{SB}$ the coupling between the two. The exact time
evolution of the reduced density matrix (RDM), $\sigma\left(t\right)={\rm Tr}_{{\rm B}}\left\{ \rho\left(t\right)\right\} $
($\rho\left(t\right)$ is the \emph{full} density matrix), within
the TCL approach, is given in terms of a time-local kernel,\citep{Tokuyama1976statistical}
$\frac{\partial}{\partial t}\sigma\left(t\right)={\cal K}\left(t\right)\sigma\left(t\right)$,
where we assumed a non-correlated initial state, namely that $\rho\left(0\right)=\sigma\left(0\right)\otimes\rho_{B}\left(0\right)$.
A simple approach to obtain ${\cal K}\left(t\right)$, is based on
the reduced system propagator, ${\cal U}_{S}\left(t\right)={\rm Tr}_{{\rm B}}\left\{ e^{\mathcal{L}t}\rho_{B}\right\} \neq e^{{\cal L}_{S}t}$,
where ${\cal L}=-\frac{i}{\hbar}\left[{\cal H},\cdots\right]$. Using
${\cal U}_{S}\left(t\right)$, the TCL generator is given by:\citep{Laird1991,Kidon2015TCL}
\begin{equation}
{\cal K}\left(t\right)=\dot{{\cal U}}_{S}\left(t\right){\cal U}_{S}^{-1}\left(t\right)\label{eq:K=00003DdU*U-1}
\end{equation}
The matrix elements of the super-operator ${\cal U}_{S}\left(t\right)$
can be obtained directly from the reduced density matrix elements,
$\sigma\left(t\right)$ evolved from different initial conditions
of the system.\citep{Kidon2015TCL} Since the time evolution of the
reduced density operator in matrix form reads 
\begin{equation}
\sigma_{ij}\left(t\right)\sideset{=}{_{kl}}\sum{\cal U}_{S,ij,kl}\left(t\right)\sigma_{kl}\left(0\right),\label{eq:sigma(t)=00003D00003DU(t)sigma(0)}
\end{equation}
it clearly follows that an initial state with $\sigma_{mm}\left(0\right)=1$
and the remaining values of $\sigma_{kl}\left(0\right)=0$, will give
${\cal U}_{S,ij,mm}\left(t\right)=\sigma_{ij}\left(t\right)$. For
more details see Ref.~\onlinecite{Kidon2015TCL}. Importantly, the
relation between ${\cal U}_{S}\left(t\right)$ and $\sigma\left(t\right)$
holds for any system-bath Hamiltonian and thus, simplifies the calculation
of ${\cal K}\left(t\right)$ for complex model systems. However, Eq.~(\ref{eq:K=00003DdU*U-1})
is ill-defined when ${\cal U}_{S}\left(t\right)$ is singular (for
example, when two system states are degenerate and couple to the bath
in the same way), limiting its applicability.

An alternative to the TCL approach, describes the time evolution of
the reduced density matrix using a non-local memory term,\citep{nakajima_quantum_1958,zwanzig_ensemble_1960,mori_transport_1965}
$\frac{\partial}{\partial t}\sigma\left(t\right)=\mathcal{L}_{S}\sigma\left(t\right)+\frac{1}{\hbar^{2}}\int_{0}^{t}d\tau\kappa\left(\tau\right)\sigma\left(t-\tau\right)$.
For this approach (TC), it is well known that one can rewrite the
memory kernel $\kappa\left(t\right)$ in terms of a Volterra equation
of the second kind:\citep{Geva2003,zhang_nonequilibrium_2006} 
\begin{equation}
\kappa\left(t\right)=\frac{\partial\Phi\left(t\right)}{\partial t}-\Phi\left(t\right)\mathcal{L}_{S}-\frac{1}{\hbar^{2}}\int_{0}^{t}d\tau\,\Phi\left(t-\tau\right)\kappa\left(\tau\right),\label{eq:volterra}
\end{equation}
where $\Phi\left(t\right)=\hbar^{2}{\rm Tr}_{{\rm B}}\left\{ \mathcal{L}e^{\mathcal{L}t}\rho_{B}\right\} $
is a super-operator that can be calculated by a variety of solvers.\citep{Makri1995,muhlbacher_real-time_2008,weiss_iterative_2008,Wang2009,gull10_bold_monte_carlo,Segal10,Cohen-Gull-Reichman2015-introducing-inchworm}
However, unlike the reduced propagator, $\Phi\left(t\right)$ depends
explicitly on the form of the system-bath coupling via the full Liouvillian
${\cal L}={\cal L}_{S}+{\cal L}_{B}+{\cal L}_{SB}$, and requires
the calculation of high--order observables in the system and bath
degrees of freedom, not required for ${\cal U}_{S}\left(t\right)$.
Here we show, the need to calculate higher--order observables can
be removed by noting a simple connection between $\Phi\left(t\right)$
and ${\cal U}_{S}\left(t\right)$: 
\begin{equation}
\frac{\Phi\left(t\right)}{\hbar^{2}}={\rm Tr}_{{\rm B}}\left\{ \mathcal{L}e^{\mathcal{L}t}\rho_{B}\right\} =\frac{\partial}{\partial t}{\rm Tr}_{{\rm B}}\left\{ e^{\mathcal{L}t}\rho_{B}\right\} =\frac{\partial{\cal U}_{S}\left(t\right)}{\partial t}.
\end{equation}
The equation for the TC memory kernel may now be rewritten in terms
of the reduced system propagator and the system Liouvillian alone:
\begin{equation}
\kappa\left(t\right)=\hbar^{2}\ddot{{\cal U}}_{S}\left(t\right)-\hbar^{2}\dot{{\cal U}}_{S}\left(t\right)\mathcal{L}_{S}-\int_{0}^{t}d\tau\,\dot{{\cal U}}_{S}\left(t-\tau\right)\kappa\left(\tau\right).\label{eq:volterra-1}
\end{equation}

Eq.~(\ref{eq:volterra-1}) is the main result of this note. It provides
a scheme to calculate the memory kernel for a general form of the
coupling Hamiltonian without the need to calculate any terms involving
the bath operators, as long as the time derivatives of ${\cal U}_{S}\left(t\right)$
are obtained numerically. We demonstrate this for the generalized
Anderson-Holstein model describing an impurity with on-site electron-electron
(e-e) interactions, coupled to three baths: a phonon bath and two
fermionic baths (leads) at different chemical potentials (full description
given in Ref.~\onlinecite{Thoss2013}). We deployed the multilayer
multiconfiguration time-dependent Hartree (ML-MCTDH) method~\citep{Thoss03}
to numerically compute the RDM at short times from independent initial
system states. The reduced propagator was obtained from the ML-MCTDH
results according to Eq.~(\ref{eq:sigma(t)=00003D00003DU(t)sigma(0)}),
and its time derivatives performed numerically ($5$ points finite
difference). Finally, the memory kernel was computed according to
Eq.~(\ref{eq:volterra-1}).

\begin{figure}[h]
\medskip{}
 \includegraphics[width=0.9\columnwidth]{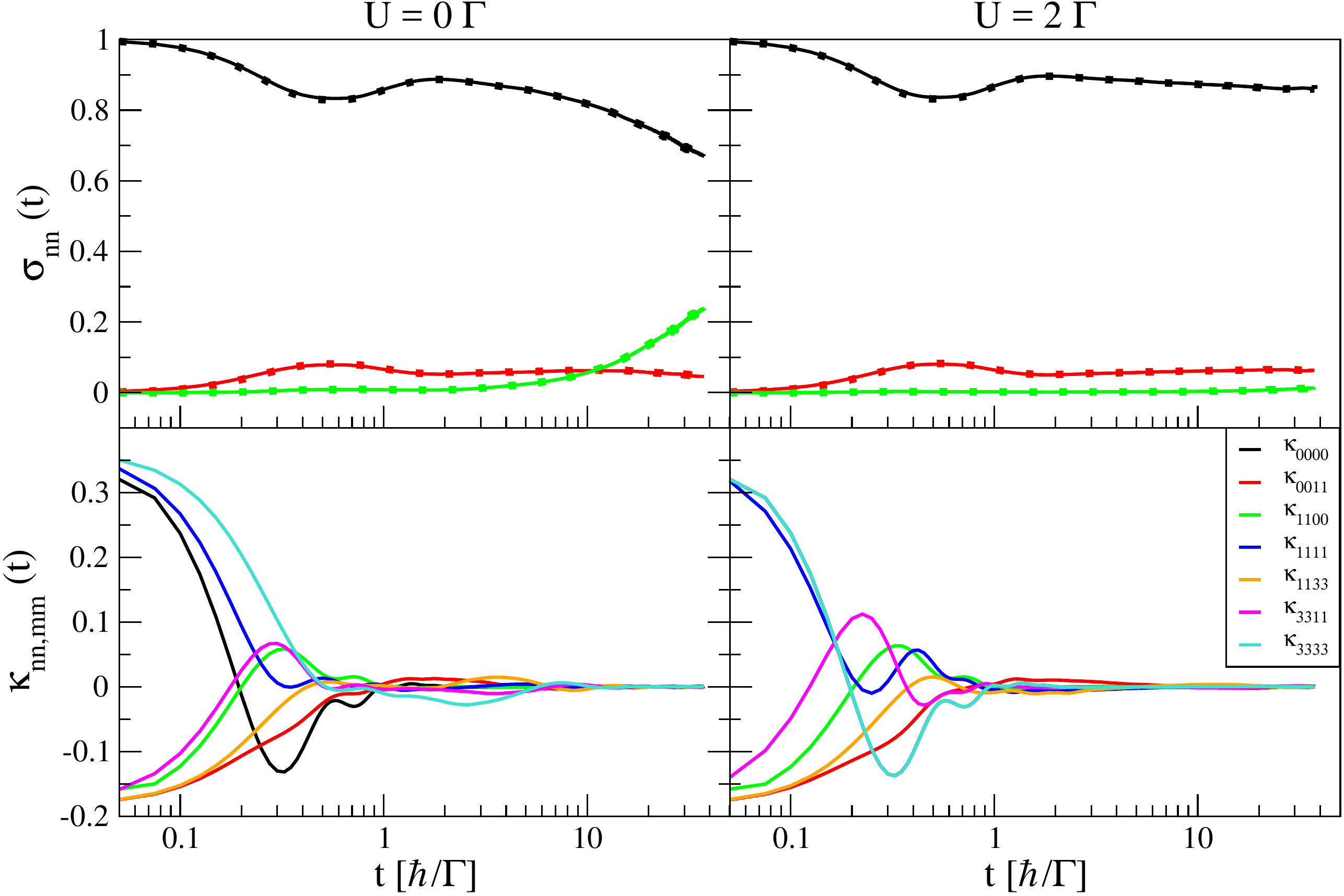}\caption{\label{fig:The-reduced-density}The RDM and memory kernel for two
values of the interaction energy $U$. Upper panels: The RDM elements
propagated from an initially empty dot using the ML-MCTDH method (squares)
and the TC GQME approach (solid lines). $\sigma_{nn}$ are the probabilities
of the dot being empty (black), occupied by one electron (red), and
occupied by two electrons (green). Lower panels: The seven distinct
nonzero memory-kernel elements. All quantities are shown in units
of the system-leads coupling strength $\Gamma$.}
\end{figure}

In Fig.~\ref{fig:The-reduced-density} we shown the results for two
values of the on-site e-e repulsion, $U$. The elements of the memory
computed from the reduced system propagator are shown in the lower
panels and the resulting elements of the reduced density are shown
in the upper panels. The populations obtained by solving the TC GQME
with the memory kernel given by Eq.~(\ref{eq:volterra-1}) are in
excellent agreement with the numerical results obtained directly from
the ML-MCTDH method (upper panels), reassuring the numerical procedure
to obtain $\kappa\left(t\right)$ from ${\cal U}_{S}\left(t\right)$.

In summary, we have related the memory kernel $\kappa\left(t\right)$
in the Nakajima--Zwanzig--Mori TC formalism to the reduced system
propagator ${\cal U}_{S}\left(t\right)$, which can be obtained at
short times from an impurity solver. Compared to previous formulations
our approach provides a robust and simpler framework, circumventing
the need to compute high-order system-bath observables. Moreover,
unlike the Tokuyama--Mori TCL approach, the current formalism does
not rely on the inversion of a super-operator which can be singular.
We illustrated the correctness of the proposed approach for a model
system describing both electron-electron and electron-phonon correlations
and find excellent agreement between the accurate ML-MCTDH results
and the generalized quantum master equation.

\def\bibfont{\footnotesize}\bibliographystyle{aipnum4-1}
\bibliography{memory-from-reduced-propagator}

\end{document}